\documentclass
[aps,preprint,tightenlines,superscriptaddress,bibnotes,titlepage,balancelastpage,secnumarabic]{revtex4}%
\usepackage{amsfonts}
\usepackage{amsmath}
\usepackage{amssymb}
\usepackage{graphicx}%
\begin{document}
\title[ ]{Chaotic Maps, Hamiltonian Flows, and Holographic Methods}
\author{Thomas L. Curtright}
\affiliation{Department of Physics, University of Miami, Coral Gables, FL 33124-8046, USA}
\author{Cosmas K. Zachos}
\affiliation{High Energy Physics Division, Argonne National Laboratory, Argonne, IL
60439-4815, USA\medskip\medskip}

\begin{abstract}
Holographic functional methods are introduced as probes of discrete
time-stepped maps that lead to chaotic behavior. \ The methods provide
continuous time interpolation between the time steps, thereby revealing the
maps to be quasi-Hamiltonian systems underlain by novel potentials that govern
the motion of a perceived point particle. \ Between turning points, the
particle is strictly driven by Hamiltonian dynamics, but at each encounter
with a turning point the potential changes abruptly, loosely analogous to the
switchbacks on a mountain road. \ A sequence of successively deepening
switchback potentials explains, in physical terms, the frequency cascade and
trajectory folding that occur on the particular route to chaos revealed by the
logistic map.

\end{abstract}
\volumeyear{year}
\volumenumber{number}
\issuenumber{number}
\eid{identifier}
\startpage{1}
\endpage{ }
\maketitle

\section{Introduction}

In a previous paper \cite{CZ}\ we have discussed how functions of position,
defined on a discrete lattice of time points, may be analytically interpolated
in time by functions of both $x$ and $t$, defined on a continuum of time
points, framed by the lattice, through the use of solutions to Schr\"{o}der's
nonlinear functional equation \cite{S}. \ For a more pedagogical discussion to
augment this Introduction, we encourage readers to look at that companion paper.

If the effect of the first discrete time step is given as the map $x\mapsto
f_{1}\left(  x,s\right)  $, for some parameter $s$, Schr\"{o}der's functional
equation is%
\begin{equation}
s\Psi\left(  x,s\right)  =\Psi\left(  f_{1}\left(  x,s\right)  ,s\right)  \ ,
\label{SE}%
\end{equation}
with $\Psi$ to be determined. \ So,
\begin{equation}
f_{1}\left(  x,s\right)  =\Psi^{-1}\left(  s\Psi\left(  x,s\right)  ,s\right)
\ ,
\end{equation}
where the inverse function $\Psi^{-1}$ obeys Poincar\'{e}'s equation,%
\begin{equation}
\Psi^{-1}\left(  sx,s\right)  =f_{1}\left(  \Psi^{-1}\left(  x,s\right)
,s\right)  \ . \label{PE}%
\end{equation}
The $n$th iterate of (\ref{SE}) gives
\begin{equation}
s^{n}\Psi\left(  x,s\right)  =\Psi\left(  f_{1}\left(  \cdots f_{1}\left(
f_{1}\left(  x,s\right)  ,s\right)  \cdots,s\right)  ,s\right)  \ ,
\end{equation}
with $f_{1}$ acting $n$ times, and thus the $n$th order functional composition
--- the so-called \textquotedblleft splinter\textquotedblright\ of the
functional equation ---%
\begin{equation}
f_{n}\left(  x,s\right)  \equiv f_{1}\left(  \cdots f_{1}\left(  f_{1}\left(
x,s\right)  ,s\right)  \cdots,s\right)  =\Psi^{-1}\left(  s^{n}\Psi\left(
x,s\right)  ,s\right)  \ .
\end{equation}
A continuous interpolation between the integer lattice of time points is then,
for \emph{any} $t$,%
\begin{equation}
f_{t}\left(  x,s\right)  =\Psi^{-1}\left(  s^{t}\Psi\left(  x,s\right)
,s\right)  \ . \label{Interpolation}%
\end{equation}
This can be a well-behaved and single-valued function of $x$ and $t$ provided
that $\Psi^{-1}\left(  x,s\right)  $ is a well-behaved, single-valued function
of $x$,\ even though $\Psi\left(  x,s\right)  $ might be, and typically is,
multi-valued. \ 

As discussed in \cite{CZ}, the interpolation can be envisioned as the
trajectory of a particle,
\begin{equation}
x\left(  t\right)  =f_{t}\left(  x,s\right)  \ , \label{Trajectory}%
\end{equation}
where the particle is moving under the influence of a potential according to
Hamiltonian dynamics.\ \ The velocity of the particle is then found by
differentiating (\ref{Interpolation}) with respect to $t$,\
\begin{equation}
\frac{dx\left(  t\right)  }{dt}=\left(  \ln s\right)  s^{t}\Psi\left(
x\right)  \left(  \Psi^{-1}\left(  s^{t}\Psi\left(  x\right)  \right)
\right)  ^{\prime}\ ,
\end{equation}
where any dependence of $\Psi$ on $s$ is implicitly understood. \ Therefore,
the velocity will inherit and exhibit any multi-valuedness possessed by
$\Psi\left(  x\right)  $. \ 

Indeed, suppose that Hamiltonian dynamics have been specified, trajectories
have been computed, and $f_{1}=x\left(  t=1\right)  $ has emerged\ in terms of
initial velocity and initial position $x\left(  t=0\right)  \equiv x$. \ A
solution of the functional equation (\ref{SE}) can then be constructed, and
expressed in very physical terms, as just an exponential of \emph{the time
elapsed along any such particle trajectory} that passes through position $x$.
\ That solution is%
\begin{equation}
\Psi\left(  x\right)  =s^{T\left(  x\right)  }\Psi_{0}\ ,\ \ \ T\left(
x\right)  =\int^{x}\frac{dy}{v\left(  y\right)  }\ . \label{GeneralPsi}%
\end{equation}
Here $v$ is the velocity as a function of the position along the trajectory.
\ (Hence $v\left(  x\left(  t\right)  \right)  $, if you wish. \ Of course,
for a Hamiltonian system with time translational invariance, we may supplant
$x\left(  t\right)  \rightarrow x\left(  t+\Delta t\right)  $ in all these
expressions, as well as those to follow.) \ For a particle moving in a
potential $V\left(  x\right)  $ at fixed energy, $E$, the velocity can be
expressed in the usual way as $v\left(  x\right)  =\pm\sqrt{E-V\left(
x\right)  }$, with suitably chosen mass units. \ 

When written in a form more closely related to Schr\"{o}der's functional
equation,\ the solution (\ref{GeneralPsi}) is simply%
\begin{equation}
\Psi\left(  f_{t}\left(  x\right)  \right)  =s^{\int_{x}^{f_{t}\left(
x\right)  }\frac{dy}{v\left(  y\right)  }}~\Psi\left(  x\right)  \ .
\end{equation}
At $t=1$, $\int_{x}^{f_{1}\left(  x\right)  }\frac{dy}{v\left(  y\right)  }%
=1$, and Schr\"{o}der's equation (\ref{SE}) re-emerges. \ But, as this
construction clearly shows, one must be careful at turning points where $v=0$,
especially if these are encountered at finite times along the trajectory.
\ Typically these turning points produce branch points in $\Psi$ so that it is
multi-valued. \ 

An interesting physical effect of such branch points, when they are
encountered in finite times, is the possibility to switch from one branch of
the underlying analytic potential function to another, thereby changing the
functional form of the Hamiltonian governing the motion along seemingly
identical intervals of a real-valued trajectory. \ We will therefore call
these \emph{quasi}-Hamiltonian systems, and we stress that when such switches
occur this is \emph{not} standard textbook Hamiltonian dynamics. \ Further
explanation will be given below in the context of the logistic map.

The interpolation (\ref{Interpolation}) can also be viewed as a holographic
specification on the $x,t$ plane \cite{L}, determining $f_{t}\left(  x\right)
$ in the surrounded \textquotedblleft bulk\textquotedblright\ from the total
data given at the bounding times, $\left\{  x\right\}  \cup\left\{
f_{1}\left(  x\right)  \right\}  $. \ In this point of view, \emph{fixed
points} in $x$ complete the boundary data and facilitate the solution of
Schr\"{o}der's equation through power series in $x$,\ hence leading to
$V\left(  x\right)  $ \emph{which was not known a priori} \cite{CZ}. \ 

In this holographic approach, the potential first appears as a quadratic in
$\Psi/\Psi^{\prime}$,%
\begin{equation}
V\left(  x\right)  =-\left(  \ln s\right)  ^{2}\left(  \frac{\Psi\left(
x\right)  }{\Psi^{\prime}\left(  x\right)  }\right)  ^{2}\ , \label{GeneralV}%
\end{equation}
up to an additive constant, where any dependence of $\Psi$ on $s$ is again
implicit. \ The $x$ dependence of the potential,
\begin{equation}
V\left(  x\right)  \equiv-v^{2}\left(  x\right)  \ ,
\end{equation}
follows from that of the velocity profile of the interpolation, defined and
given by \cite{LogicalConnections}%
\begin{equation}
v\left(  x\right)  \equiv\left.  \frac{df_{t}\left(  x\right)  }%
{dt}\right\vert _{t=0}=\left(  \ln s\right)  \Psi\left(  x\right)  \left(
\Psi^{-1}\left(  \Psi\left(  x\right)  \right)  \right)  ^{\prime}=\frac{\ln
s}{\frac{d}{dx}\ln\Psi\left(  x\right)  }\ . \label{Generalv}%
\end{equation}
Monotonic motions between two fixed points, such as occur for the Ricker model
\cite{R}, provide the most elementary examples \cite{CZ}. \ 

However, situations where one or both of the fixed points are absent were not
fully addressed in our previous work. \ This is precisely the situation that
occurs when turning points are encountered at finite times in the particle
dynamics interpretation, and leads to an intriguing modification in the
physical picture involving the potential and its effect on the particle
trajectories. \ We consider here a specific example of such a situation
involving the well-studied logistic map of chaotic dynamics, namely
\textit{\cite{M,CE,F,K,C}},%
\begin{equation}
x\mapsto sx\left(  1-x\right)  \ . \label{LogisticMap}%
\end{equation}
Applying our functional methods to this example, the resulting interpolations
from a discrete lattice of time points to a time continuum then allow us to
appreciate analytic features of the logistic map, and to derive the governing
differential evolution laws --- indeed, subtly time-translation-invariant
Hamiltonian dynamical laws --- of the underlying physical system, hence to
obtain potentials that were \emph{not} previously known.

For the all-familiar logistic map illustrating transition to chaos, the
functional interpolation reveals that there are well-defined expressions for
the continuous time evolution of this map for \emph{all} parametric values of
the map, whether chaotic or not. \ We obtain agreement with the explicit
closed-form solutions of (\ref{SE}) for the special values of the parameters:
$s=-2,\ 2,$\ and\ $4$. \ (These explicit solutions have been known for almost
a century and a half \cite{S}.) \ \ Moreover, as indicated, from
(\ref{GeneralV})\ we can now find the potentials needed to produce these
explicit, continuously evolving trajectories in the language of Hamiltonian dynamics.

A new feature for the potentials so obtained for the $s=4$ case in particular,
and also for other values of $s>2$, is that they must change at discrete
intervals of the envisioned Hamiltonian particle's motion, to be consistent
with the evolution trajectories: \ \emph{Every time the particle hits a
turning point, the potential changes }\cite{Footnote1}. \ We therefore call
the corresponding $V$s \textquotedblleft switchback
potentials\textquotedblright\ and to echo our previous general remarks, we
will call the dynamical system as a whole \textquotedblleft
quasi-Hamiltonian.\textquotedblright\ \ For $s=4$ the switchback potentials
deepen successively and thus lead to more rapid, higher frequency motion,
resulting in the familiar chaotic behavior of the discrete $s=4$ logistic map
that they interpolate. \ The familiar frequency-cascade-and-folding behavior
of the chaotic discrete map is thus understood from --- indeed,
\emph{explained} by --- the subtle properties of the switchback potentials.

\section{The Logistic Map}

Consider in detail the logistic map (\ref{LogisticMap}) on the unit interval,
$x\in\left[  0,1\right]  $. \ Schr\"{o}der's equation for this map is \
\begin{equation}
s\Psi\left(  x,s\right)  =\Psi\left(  sx\left(  1-x\right)  ,s\right)  \ .
\label{SFE}%
\end{equation}
The inverse function satisfies the corresponding Poincar\'{e} equation,%
\begin{equation}
\Psi^{-1}\left(  sx,s\right)  =s\Psi^{-1}\left(  x,s\right)  \left(
1-\Psi^{-1}\left(  x,s\right)  \right)  \ . \label{PFE}%
\end{equation}
As originally obtained by Schr\"{o}der, there are three closed-form solutions
known, for $s=-2,$ $2,$ and $4$:%
\begin{align}
\Psi\left(  x,-2\right)   &  =\frac{\sqrt{3}}{6}\left(  2\pi-3\arccos\left(
x-\frac{1}{2}\right)  \right)  \ ,\ \ \ \Psi^{-1}\left(  x,-2\right)
=\frac{1}{2}-\cos\left(  \frac{2x}{\sqrt{3}}+\frac{\pi}{3}\right)
\ ,\nonumber\\
\Psi\left(  x,2\right)   &  =-\frac{1}{2}\ln\left(  1-2x\right)
\ ,\ \ \ \Psi^{-1}\left(  x,2\right)  =\frac{1}{2}\left(  1-e^{-2x}\right)
\ ,\nonumber\\
\Psi\left(  x,4\right)   &  =\left(  \arcsin\sqrt{x}\right)  ^{2}%
\ ,\ \ \ \Psi^{-1}\left(  x,4\right)  =\left(  \sin\sqrt{x}\right)  ^{2}\ .
\label{ExactCases}%
\end{align}
Note that, while the $\Psi$ are multi-valued, the inverse functions are all
single-valued. \ 

More generally, consider a power series for any $s$.%
\begin{equation}
\Psi^{-1}\left(  x,s\right)  =x+x\sum_{n=1}^{\infty}x^{n}c_{n}\left(
s\right)  \ . \label{PsiInverseSeries}%
\end{equation}
The Poincar\'{e} equation then leads to a recursion relation for the
$s$-dependent coefficients.%
\begin{equation}
c_{n+1}=\frac{1}{1-s^{n+1}}\sum_{j=0}^{n}c_{j}c_{n-j}\ ,
\label{PsiInverseSeriesCoefs}%
\end{equation}
with $c_{0}=1$, $c_{1}=\frac{1}{1-s}$, $c_{2}=\frac{2}{\left(  1-s\right)
\left(  1-s^{2}\right)  }$, etc. \ The explicit coefficients are easily
recognized for $s=-2,$ $2,$ and $4$, and immediately yield the three
closed-form cases. \ Similarly let
\begin{equation}
\Psi\left(  x,s\right)  =x+x\sum_{n=1}^{\infty}\left(  -x\right)  ^{n}%
d_{n}\left(  s\right)  \ . \label{PsiSeries}%
\end{equation}
Then, as a consequence of Schr\"{o}der's equation, $d_{1}=1/\left(
1-s\right)  $, and for $n\geq2$,%
\begin{equation}
d_{n}=\frac{1}{1-s^{n}}\sum_{k=1}^{\left\lfloor \frac{n+1}{2}\right\rfloor
}\binom{n+1-k}{k}s^{n-k}d_{n-k}\ . \label{PsiSeriesCoefs}%
\end{equation}
where $\left\lfloor \cdots\right\rfloor $ is the (integer-valued) floor
function. \ In principle, these series solve (\ref{PFE}) and (\ref{SFE})\ for
any $s$, within their radii of convergence. \ 

From extensive numerical studies we believe the radius of convergence for
(\ref{PsiSeries})\ depends on $s$ as follows.%
\begin{equation}
R_{\Psi}=\left\{
\begin{array}
[c]{ccc}%
\left\vert 1-\dfrac{1}{s}\right\vert  & \text{if} & 0\leq s\leq2\ ,\\
&  & \\
\dfrac{s}{4} & \text{if} & 2\leq s\leq4\ .
\end{array}
\right.  \label{PsiSeriesRadius}%
\end{equation}
Preliminary work further suggests that these conjectural results can actually
be established analytically by comparison with known convergent series, at
least in some cases.

The radius of convergence for the inverse function series
(\ref{PsiInverseSeries}) is intriguing for $0\leq s\leq1$. \ A fit by
Mathematica to radii obtained numerically from explicit series coefficients
$c_{n}\left(  s\right)  $ up to $n=200$, for 20 different values of $s$ in
this range, using the functional form $\frac{1}{4}\left(  1-s\right)  ^{a}\exp
bs$, gives $a=0.78$ and $b=0.62$. \ Thus $R_{\Psi^{-1}}\left(  s\right)  $
falls from $1/4$ to zero as $s$ goes from $0$ to $1$. \ However, for the
parameters of most interest, namely $2\leq s\leq4$, numerical studies suggest
the series all have \emph{infinite} radius of convergence, indicating the
$\Psi^{-1}$ are entire functions for these parameters. \ This is true for
$s=2$ and $s=4$, of course, as is evident from (\ref{ExactCases}). \ The
numerics provide compelling evidence that this is also true for all
intermediate $s$.

Given convergent series, the functional equations can then be used to continue
the series solutions outside their radii of convergence and thereby obtain
accurate determinations of the various branches of $\Psi$, and of the
large-but-finite argument behavior of $\Psi^{-1}$, in a manner familiar from,
say, the $\Gamma$ and $\zeta$ functions. \ (See \cite{CV}\ for details, and
the Appendix for a numerical example.) \ Comparison of the results from these
series-plus-functional-equation methods with the known exact solutions
(\ref{ExactCases}) shows perfect agreement.

The first few terms for $\Psi$ and $\Psi^{-1}$ for generic $s$ are given
explicitly by \cite{s=0}%
\begin{gather}
\Psi\left(  x,s\right)  =x+\frac{x^{2}}{s-1}+\frac{2s}{\left(  s+1\right)
}\frac{x^{3}}{\left(  s-1\right)  ^{2}}+\frac{s\left(  1+5s^{2}\right)
}{\left(  s^{2}+s+1\right)  \left(  s+1\right)  }\frac{x^{4}}{\left(
s-1\right)  ^{3}}\nonumber\\
+\frac{2s^{3}\left(  3+2s+7s^{3}\right)  }{\left(  s^{2}+1\right)  \left(
s^{2}+s+1\right)  \left(  s+1\right)  ^{2}}\frac{x^{5}}{\left(  s-1\right)
^{4}}\nonumber\\
+\frac{2s^{3}\left(  1+3s+14s^{3}+14s^{4}+7s^{5}+21s^{7}\right)  }{\left(
s^{4}+s^{3}+s^{2}+s+1\right)  \left(  s^{2}+1\right)  \left(  s^{2}%
+s+1\right)  \left(  s+1\right)  ^{2}}\frac{x^{6}}{\left(  s-1\right)  ^{5}%
}+O\left(  x^{7}\right)  \ ,\\
\Psi^{-1}\left(  x,s\right)  =x+\frac{x^{2}}{1-s}+\frac{2}{\left(  s+1\right)
}\frac{x^{3}}{\left(  s-1\right)  ^{2}}+\frac{5+s}{\left(  s+1\right)  \left(
s^{2}+s+1\right)  }\frac{x^{4}}{\left(  1-s\right)  ^{3}}\nonumber\\
+\frac{2\left(  7+3s+2s^{2}\right)  }{\left(  s^{2}+1\right)  \left(
s^{2}+s+1\right)  \left(  s+1\right)  ^{2}}\frac{x^{5}}{\left(  s-1\right)
^{4}}\nonumber\\
+\frac{2\left(  21+14s+14s^{2}+8s^{3}+3s^{4}\right)  }{\left(  s^{4}%
+s^{3}+s^{2}+s+1\right)  \left(  s^{2}+1\right)  \left(  s^{2}+s+1\right)
\left(  s+1\right)  ^{2}}\frac{x^{6}}{\left(  1-s\right)  ^{5}}+O\left(
x^{7}\right)  \ ,
\end{gather}
from which we infer that $\Psi\left(  x,s\right)  /x$ and $\Psi^{-1}\left(
x,s\right)  /x$ are actually series in $x/\left(  1-s\right)  $ with
$s$-dependent coefficients that are analytic near $s=1$.

The trajectories interpolating the splinter (integer $t$) of the logistic map
are then,%
\begin{gather}
x\left(  t\right)  =\Psi^{-1}\left(  s^{t}\Psi\left(  x,s\right)  ,s\right)
=s^{t}~x+\frac{s^{t}\left(  1-s^{t}\right)  }{s-1}~x^{2}+\frac{2s^{t}\left(
1-s^{t}\right)  \left(  s-s^{t}\right)  }{\left(  s+1\right)  \left(
s-1\right)  ^{2}}~x^{3}\nonumber\\
+\frac{s^{t}\left(  1-s^{t}\right)  \left(  s-s^{t}\right)  \left(
1+5s^{2}-\left(  s+5\right)  s^{t}\right)  }{\left(  s+1\right)  \left(
s^{2}+s+1\right)  \left(  s-1\right)  ^{3}}~x^{4}\nonumber\\
+\frac{2s^{t}\left(  1-s^{t}\right)  \left(  s-s^{t}\right)  \left(
s^{2}-s^{t}\right)  \left(  7s^{3}+2s+3-s^{t}\left(  2s^{2}+3s+7\right)
\right)  }{\left(  s+1\right)  ^{2}\left(  s^{2}+1\right)  \left(
s^{2}+s+1\right)  \left(  s-1\right)  ^{4}}~x^{5}+O\left(  x^{6}\right)  \ .
\end{gather}
The trajectories are single-valued functions of the time so long as $\Psi
^{-1}$ is single-valued, and in fact, they exist even for $s\rightarrow1$ as
formal series solutions. $\ $Explicitly,
\begin{gather}
\lim\limits_{s\rightarrow1}\Psi^{-1}\left(  s^{t}\Psi\left(  x,s\right)
,s\right)  =x-t~x^{2}+t\left(  t-1\right)  ~x^{3}-\frac{1}{2}t\left(
t-1\right)  \left(  2t-3\right)  ~x^{4}\nonumber\\
+\frac{1}{3}t\left(  t-1\right)  \left(  t-2\right)  \left(  3t-4\right)
~x^{5}-\frac{1}{12}t\left(  t-1\right)  \left(  t-2\right)  \left(
12t^{2}-41t+31\right)  ~x^{6}\nonumber\\
+\frac{1}{30}t\left(  t-1\right)  \left(  t-2\right)  \left(  30t^{3}%
-171t^{2}+302t-157\right)  ~x^{7}+O\left(  x^{8}\right)  \ .
\end{gather}

For the three special cases,\ $s=-2,\ 2,$ and $4$, there are closed-form
results for various quantities of interest. \ For example, for $s=4$, the
trajectory and velocity are given by%
\begin{equation}
\left.  x\left(  t\right)  \right\vert _{s=4}=\Psi^{-1}\left(  4^{t}%
~\Psi\left(  x,4\right)  ,4\right)  =\left(  \sin\left(  2^{t}\arcsin\sqrt
{x}\right)  \right)  ^{2}\ , \label{ExactTrajectory4}%
\end{equation}
(as originally presented in \cite{S}, p 306) and by%
\begin{subequations}
\begin{align}
\frac{d}{dt}x\left(  t\right)   &  =\left(  2^{1+t}\ln2\right)  \sin\left(
2^{t}\arcsin\sqrt{x}\right)  \cos\left(  2^{t}\arcsin\sqrt{x}\right)
\arcsin\sqrt{x}\\
&  =\left(  \ln4\right)  \sqrt{x\left(  t\right)  \left(  1-x\left(  t\right)
\right)  }\arcsin\sqrt{x\left(  t\right)  }\ .
\end{align}
The last expression evinces a continuous time-translational invariance.
\ However, this velocity function has branch points (i.e. turning points) at
$x\left(  t\right)  =0$ and $x\left(  t\right)  =1$, so some care is needed to
determine which branch of the function is involved, particularly when the
turning points are encountered at finite times. \ The system is therefore a
quasi-Hamiltonian one, as we have previously indicated, and as we shall
discuss in more detail.

With this caveat in mind, we may thus deduce the velocity profile $v\left(
x\right)  =\left.  \frac{dx\left(  t\right)  }{dt}\right\vert _{t=0}$, the
effective potential, and the force for the underlying quasi-Hamiltonian
system:%
\end{subequations}
\begin{align}
v\left(  x\right)   &  =\left.  \frac{dx\left(  t\right)  }{dt}\right\vert
_{s=4,t=0}=\left(  \ln4\right)  \sqrt{x\left(  1-x\right)  }\arcsin\sqrt
{x}\ ,\\
V\left(  x\right)   &  =-v^{2}\left(  x\right)  =\left(  \ln4\right)
^{2}x\left(  x-1\right)  \arcsin^{2}\sqrt{x}\ ,\\
F\left(  x\right)   &  =-\frac{d}{dx}V\left(  x\right)  =\left(  \ln4\right)
^{2}\left(  \arcsin\sqrt{x}\right)  \left(  \sqrt{x\left(  1-x\right)
}-\left(  2x-1\right)  \arcsin\sqrt{x}\right)  \ .
\end{align}
Note $x=0$ is an unstable fixed point for the system. \ Also, the system is
time-translationally invariant, so these same expressions hold at all times
everywhere along a trajectory, \emph{provided} due care is taken to determine
which branch of the various functions is in effect. Thus $V\left(  x\left(
t\right)  \right)  =\left(  \ln4\right)  ^{2}x\left(  t\right)  \left(
x\left(  t\right)  -1\right)  \arcsin^{2}\sqrt{x\left(  t\right)  }$, etc.

Similar closed-form results hold for the other two solvable cases, $s=-2$ and
$s=2$. \ In principle, there are also potentials underlying the trajectories
for other $s$ by dint of the series constructions. \ (We illustrate one other
case, $s=3$, in the Appendix.)

From the general expression for $x\left(  t\right)  $ in terms of $\Psi$ and
$\Psi^{-1}$, and with use of the chain rule, the potential may be expressed
entirely in terms of $\left(  \ln\Psi\right)  ^{\prime}$, as given\ in the
Introduction by (\ref{GeneralV}). \ The Schr\"{o}der auxiliary function is
then recognized as just an exponential of the time function
\begin{equation}
T\left(  x\right)  =\int^{x}\frac{dy}{\sqrt{-V\left(  y\right)  }}\ ,
\end{equation}
computed along a zero-energy trajectory, as given by (\ref{GeneralPsi}).

\section{Novel potentials and switchback effects}

The effective potentials for all three closed-form cases are somewhat unusual:
\ \
\begin{align}
V\left(  x,s=4\right)   &  =\left(  \ln4\right)  ^{2}x\left(  x-1\right)
\arcsin^{2}\sqrt{x}\ ,\\
V\left(  x,s=2\right)   &  =-\left(  \ln\sqrt{2}\right)  ^{2}\left(
1-2x\right)  ^{2}\ln^{2}\left(  1-2x\right)  \ ,\\
V\left(  x,s=-2\right)   &  =\frac{1}{36}\left(  \ln\left(  -2\right)
\right)  ^{2}\left(  2x+1\right)  \left(  2x-3\right)  \left(  2\pi
-3\arccos\left(  x-\frac{1}{2}\right)  \right)  ^{2}\ .
\end{align}
Another way to express the potential for $s=4$ is similar in form to that for
$s=-2$, namely,%
\begin{equation}
V\left(  x,s=4\right)  =\left(  \ln2\right)  ^{2}x\left(  x-1\right)  \left(
\pi-\arccos\left(  2x-1\right)  \right)  ^{2}\ .
\end{equation}
Indeed, it is well-known that the logistic maps for $s=4$ and $s=-2$ are
intimately related through the functional conjugacy of the underlying
Schr\"{o}der equations. \ But note the potential for $s=-2$ is in fact
complex, since $\ln\left(  -2\right)  =\ln2+i\pi$, as are the trajectories
under real time evolution for this case. \ Of course, since the complexity of
$V\left(  x,s=-2\right)  $\ is solely a multiplicative factor, if we switch to
complex time, $\tau=\left(  \ln2+i\pi\right)  \times t$, then%
\begin{equation}
\frac{d^{2}x}{d\tau^{2}}=-\frac{\partial}{\partial x}\frac{V\left(
x,s=-2\right)  }{2\left(  \ln2+i\pi\right)  ^{2}}%
\end{equation}
again yields real trajectories $x\left(  \tau\right)  $. \ As trivial as this
observation is, nevertheless it does raise several questions about complex
$x$, and about the behavior of $V$ in the complex plane. \ We discuss only one
aspect of this behavior here, and consider non-principal values for the
multi-valued functions $v\left(  x,s=4\right)  $ and $V\left(  x,s=4\right)
$. \ In particular, \emph{all} branches of the $\arcsin$ function are
important to understand the behavior of the explicit trajectories
(\ref{ExactTrajectory4}). \ 

\textbf{Switchbacks on the road to chaos. \ }An interesting \emph{new}
feature, which we shall call the \emph{switchback effect}, appears for a
particle moving in the $s=4$ effective potential. \ This is a distinguishing
feature that we encounter for the $s=4$ chaotic map, but not, say, for the
non-chaotic $s=2$ map. \ The effect is traceable to the fact that, while the
trajectory is single-valued as a function of the time, the velocity and hence
the effective potential are \emph{not} single-valued as functions of
\emph{position}. \ Moreover, the branch points of the multi-valued functions
are encountered by zero-energy trajectories at \emph{finite} times for the
$s=4$ case, unlike, say, for the $s=2$ case. \ For the latter case, it takes
an infinite time for a particle with zero energy to reach a turning point.

Switchbacks are essentially transitions from one branch of the
position-dependent velocity function to another, and occur at the turning
points encountered at finite times by the interpolating particle trajectories,
(\ref{ExactTrajectory4}). \ The effect is easily seen upon viewing animations
of these trajectories \cite{Animations}.

Consider a particle with zero total energy that starts in the $V\left(
x,s=4\right)  $ potential given above, in the region $0<x<1$. \ If the
particle is initially moving to the left, it will continue towards $x=0$,
taking an infinite amount of time to reach that turning point. \ But if the
particle is initially moving to the right, it will reach the $x=1$ turning
point in a \emph{finite }amount of time that depends on its initial $x$, as
given by
\begin{equation}
\Delta t_{0}\left(  x\right)  =\frac{1}{\ln4}\int_{x}^{1}\frac{dy}%
{\sqrt{y\left(  1-y\right)  }\left(  \arcsin\sqrt{y}\right)  }=\frac{1}{\ln
2}~\ln\left(  \frac{\pi/2}{\arcsin\sqrt{x}}\right)  \ .
\end{equation}
So, for example, starting from the midpoint $x=1/2$, it takes unit time to
reach the turning point at $x=1$.

Upon reaching $x=1$, the explicit form of the time-dependent solution
(\ref{ExactTrajectory4}) exhibits the classical counterpart of\emph{ a sudden
transition} that keeps $E=0$, but \emph{changes the potential} for the return
trip towards $x=0$, exactly as follows from the particle moving on a different
branch of the $\arcsin$ function. \ Explicitly, the potential deepens to%
\begin{equation}
V\left(  x,s=4\right)  \Longrightarrow V_{1}\left(  x\right)  =\left(
\ln4\right)  ^{2}x\left(  x-1\right)  \left(  -\pi+\arcsin\sqrt{x}\right)
^{2}\ ,
\end{equation}
with the particle's speed changing accordingly as a function of $x$. $\ $The
return velocity profile is now negative, and given by%
\begin{equation}
v_{1}\left(  x\right)  =\left(  \ln4\right)  \sqrt{x\left(  1-x\right)
}\left(  -\pi+\arcsin\sqrt{x}\right)  \ .
\end{equation}
The $\arcsin$ in this last expression, as well as in $V_{1}$, is understood to
be the principal value.

Moving in this modified negative potential, it now takes the zero-energy
particle a \emph{finite} amount of time to travel from $x=1$ down to $x=0$, as
given by $\Delta t_{1}=1$. \ Upon reaching $x=0$, the exact solution
(\ref{ExactTrajectory4}) exhibits another sudden transition that keeps $E=0$,
but again alters the potential for the return trip towards $x=1$. \ The
potential becomes%
\begin{equation}
V_{1}\left(  x\right)  \Longrightarrow V_{2}\left(  x\right)  =\left(
\ln4\right)  ^{2}x\left(  x-1\right)  \left(  \pi+\arcsin\sqrt{x}\right)
^{2}\ ,
\end{equation}
with corresponding changes in the velocity profile. \ Again, a finite amount
of time is needed for the particle to go from $x=0$ to $x=1$ in the potential
$V_{2}$, as given by $\Delta t_{2}=\ln\left(  \frac{3}{2}\right)  /\ln2$. \ 

Upon reaching $x=1$ again, the switchback process continues. \ The total
energy remains at zero, but the potential for the second return trip further
deepens,%
\begin{equation}
V_{2}\left(  x\right)  \Longrightarrow V_{3}\left(  x\right)  =\left(
\ln4\right)  ^{2}x\left(  x-1\right)  \left(  -2\pi+\arcsin\sqrt{x}\right)
^{2}\ ,
\end{equation}
etc. \ In order for the particle to follow the interpolating trajectory
(\ref{ExactTrajectory4}) specified by the Schr\"{o}der equation for the $s=4$
case, it has to move under the influence of successively deepening potentials. \ 

At later times, the effective potential seen by the particle on its zigzag
path between the turning points will depend on the total number of previous
encounters with those turning points. \ In this sense, the particle remembers
its past. \ Let $P$ be the total number of turning points previously
encountered on the particle's trajectory. \ The motion of the particle\ before
encountering the next, $\left(  P+1\right)  $st, turning point is then
completely determined, for $s=4$, by the effective potential (again, $\arcsin$
here is understood to be the principal value)%
\begin{equation}
V_{P}\left(  x\right)  =\left(  \ln4\right)  ^{2}x\left(  x-1\right)  \left(
\left(  -\right)  ^{P}\left\lfloor \frac{1+P}{2}\right\rfloor \pi+\arcsin
\sqrt{x}\right)  ^{2}\ , \label{Indexed V}%
\end{equation}
where $\left\lfloor \cdots\right\rfloor $ is again the floor function. \ That
is to say, the potential deepens as $P$ increases. \ The corresponding
velocity profile speeds up:
\begin{equation}
v_{P}\left(  x\right)  =\left(  \ln4\right)  \sqrt{x\left(  1-x\right)
}\left(  \left(  -\right)  ^{P}\left\lfloor \frac{1+P}{2}\right\rfloor
\pi+\arcsin\sqrt{x}\right)  \ . \label{Indexed v}%
\end{equation}
The particle will either be traveling to the left, with $v_{P}\left(
x\right)  <0$ for odd $P$, or traveling to the right, with $v_{P}\left(
x\right)  >0$ for even $P$, with its speed increasing with $P$. \ As mentioned
previously, this effect is clearly seen upon viewing numerical animations of
the $s=4$ trajectories \cite{Animations}. \ It is also instructive to plot
$E\left(  t\right)  =v^{2}\left(  x\left(  t\right)  \right)  +V_{P}\left(
x\left(  t\right)  \right)  $ versus $t$ for various $P$, with $t\geq
t_{P\text{ on}}$, to check $E=0$, as well as to see how the energy would
\emph{not} be conserved if the potentials were \emph{not} switched. \ See
Figure 1.%
\begin{figure}[ptb]%
\centering
\includegraphics[
height=2.0689in,
width=5.1664in
]%
{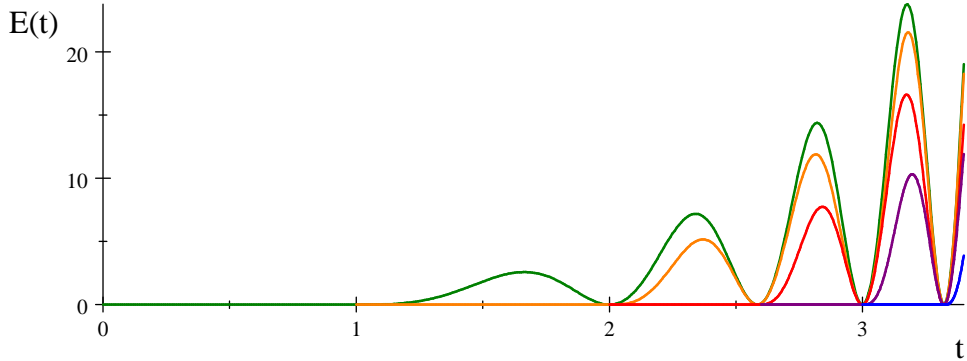}%
\caption{$E\left(  t\right)  $ for initial $x=1/2$, using potentials $V_{P}$,
with $P=0$, $1$, $2$, $3$, \& $4$.}%
\end{figure}

The time for the zero-energy particle to traverse the complete unit interval
in $x$ while moving through the $V_{P}$ potential (the $P$th \textquotedblleft
pseudo-half-cycle\textquotedblright) is always finite, for $P\neq0$:%
\begin{equation}
\Delta t_{P}=\left\vert \int_{0}^{1}\frac{dx}{v_{P}\left(  x\right)
}\right\vert =\frac{1}{\ln2}~\left\vert \ln\left(  1+\frac{1}{2\left(
-\right)  ^{P}\left\lfloor \frac{1+P}{2}\right\rfloor }\right)  \right\vert
=\frac{1}{\ln2}\ln\left(  \frac{P+1}{P}\right)  \ .
\end{equation}
This \emph{transit time}\ decreases monotonically as $P$ increases, with
$\Delta t_{P}\underset{P\rightarrow\infty}{\sim}\frac{1}{P\ln2}$. \ Starting
from an initial $x$, with initial $v>0$, the times at which changes in the
potential occur, i.e. the times at which the particle encounters turning
points, are obtained by summing these transit times and then adding $\Delta
t_{0}\left(  x\right)  $. \ The transit time sums are simple enough
:%
\begin{equation}
\sum_{N=1}^{P-1}\Delta t_{N}=\frac{\ln P}{\ln2}\ .
\end{equation}
Thus, a particle beginning at $x$, with initial $v>0$, will encounter its
$P$th turning point, and the potential $V_{P}$ will switch on, at time
$t_{P\text{ on}}\left(  x\right)  =\Delta t_{0}\left(  x\right)  +\frac{\ln
P}{\ln2}$. \ The potential $V_{P}$ will remain in effect for a time span equal
to the particle's transit time, $\Delta t_{P}=t_{P+1\text{ on}}\left(
x\right)  -t_{P\text{ on}}\left(  x\right)  $, i.e. until the next potential
$V_{P+1}$ switches on. \ 

The potential index $P$ is in fact related to the action for the
pseudo-half-cycle \cite{Morse}. \ For even $P$, we have $I\propto\int_{0}%
^{1}\cdots$, while for odd $P$, we have $I\propto\int_{1}^{0}\cdots$. \ Thus
(in the units chosen, the mass is $m=2$)%
\begin{equation}
I_{P}=2\left(  -1\right)  ^{P}\int_{0}^{1}v_{P}\left(  x\right)  dx=\left(
4\left\lfloor \frac{1+P}{2}\right\rfloor +\left(  -1\right)  ^{P}\right)
I_{0}\ , \label{PthCrossingAction}%
\end{equation}
where for the first pass through the complete unit interval the action is
\begin{equation}
I_{0}\equiv2\int_{0}^{1}v\left(  x\right)  dx=\left(  4\ln2\right)  \int
_{0}^{1}\sqrt{x}\sqrt{1-x}\arcsin\sqrt{x}dx=\frac{1}{8}\left(  \ln2\right)
\pi^{2}\ .
\end{equation}
Moreover, $2\int_{0}^{1}\left(  -v_{2k+1}\left(  x\right)  +v_{2k+2}\left(
x\right)  \right)  dx=8\left(  k+1\right)  I_{0}$ for integer $k$, so after
the initial crossing, the $\left(  k+1\right)  $st zigzag $1\rightarrow
0\rightarrow1$ (\textquotedblleft pseudo-full-cycle\textquotedblright)
contributes to the action: $\ $%
\begin{equation}
I_{k+1\ \leftrightarrows}=8\left(  k+1\right)  I_{0}\ .
\end{equation}

The progressive deepening of the potential is also evident in the expanding
phase-space trajectories for the evolving particle, where the switchbacks
enhance the velocity after each encounter with a turning point ($v=0$). \ For
example, see Figure 2.%
\begin{figure}[ptb]%
\centering
\includegraphics[
height=3.0095in,
width=4.5363in
]%
{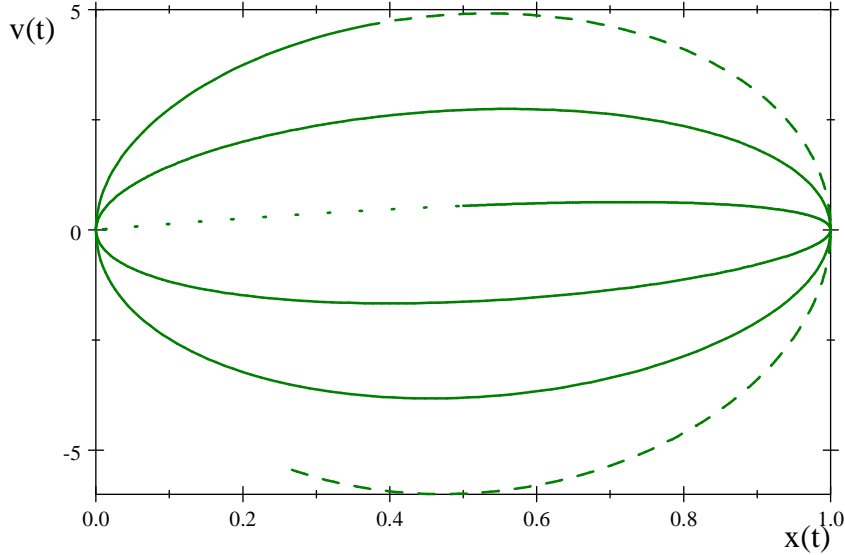}%
\caption{Phase-space trajectory $\left(  x\left(  t\right)  ,v\left(
t\right)  \right)  $ for $s=4$, with $x\left(  0\right)  =1/2$ and $v\left(
0\right)  =\frac{1}{4}\pi\ln2$. \ The solid portion of the curve is for $0\leq
t\leq3$, while the dashed extension is for $3\leq t\leq3.5$. \ The dotted
portion of the curve is for negative times, $-\infty\leq t\leq0$, with
$x\left(  -\infty\right)  =v\left(  -\infty\right)  =0$.}%
\end{figure}
This Figure provides a good opportunity to clarify the differences between
conventional Hamiltonian and quasi-Hamiltonian systems. \ The apparent failure
to be a conventional Hamiltonian system is due entirely to the fact that a
given phase-space trajectory seems to repeatedly cross itself at two points
($x\left(  t\right)  =0$, and $x\left(  t\right)  =1$, as shown in the Figure)
and at various times. Thus a vector field on the phase space needed to
describe the motion in the conventional Hamiltonian formalism would appear to
be singular at those two points. \ 

These phase-space singularities are tied to the branch point and Riemann
surface sheet structure of the relevant analytic functions. \ In fact, an
alternate way to visualize the motion is as a trajectory on the sheets of a
Riemann surface. \ Consider the particle moving on the complex $x$ plane, and
not just the real line segment $\left[  0,1\right]  $, the endpoints of which
are now branchpoints. \ There are cuts from $+1$ to $+\infty$ and from $0$ to
$-\infty$. \ The particle first moves along the real axis, approaches the cut
at $+1$, and then goes around the branchpoint such that\ $\sqrt{x}%
\rightarrow\sqrt{x}$, $\sqrt{1-x}\rightarrow-\sqrt{1-x}$, and $\arcsin
\rightarrow\pi-\arcsin$. \ The particle returns along the real axis to the
origin and encircles the branchpoint at $0$, such that $\sqrt{x}%
\rightarrow-\sqrt{x}$, $\sqrt{1-x}\rightarrow\sqrt{1-x}$, and $\arcsin\sqrt
{x}\rightarrow-\arcsin\sqrt{x}$. \ The particle then goes back to $+1$ and
goes around it once more such that again $\sqrt{x}\rightarrow\sqrt{x}$,
$\sqrt{1-x}\rightarrow-\sqrt{1-x}$, and $\arcsin\rightarrow\pi-\arcsin$. \ The
trajectory continues in this way, flipping signs and adding $\pi$s according
to the formulae (\ref{Indexed V}) and (\ref{Indexed v}) \cite{Footnote2}.

For particles with initial $v>0$, a third way to picture the dynamics is in
terms of the total distance traveled by the particle. \ In this point of view,
the successive $V_{P}$ patch together to form a continuous potential $V\left(
X\right)  $ on the real half-line, $X\geq0$, as shown in Figure 3. \ Indeed,
the half-line may be thought of as a \emph{covering manifold} of the unit
interval in $x$, with the previous multi-valued functions of $x$ now
single-valued functions of $X$. \ To avoid stagnation at any of the cusps of
$V\left(  X\right)  $, either physical \emph{or} conceptual, in this picture
the $E=0$ right-moving trajectories $X\left(  t\right)  $ may be considered as
limits of positive energy right-movers, $X\left(  t\right)  =\lim
_{E\downarrow0}X_{E}\left(  t\right)  $. \ This limiting process corresponds
to the encirclement of the branch points in the Riemann surface picture.%
\begin{figure}[ptb]%
\centering
\includegraphics[
height=3.0369in,
width=4.5645in
]%
{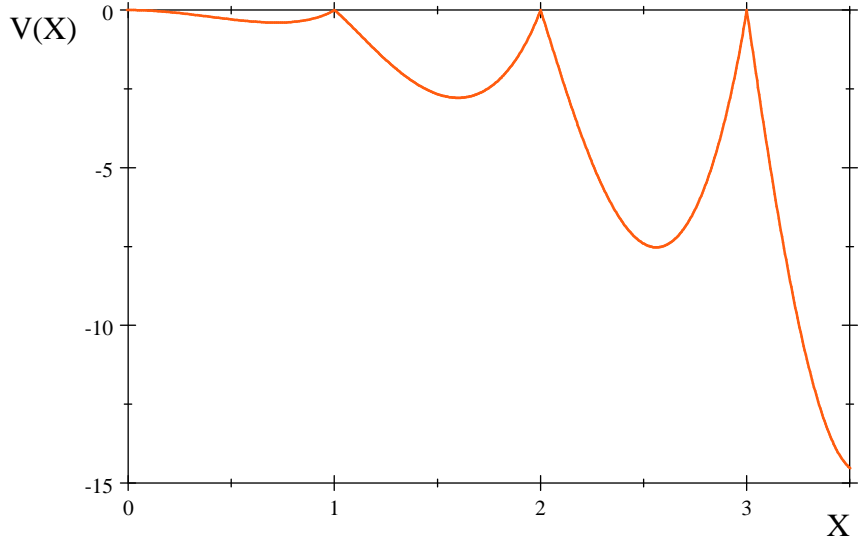}%
\caption{$V(X)$ where the total distance traveled by the particle is
$X-x\left(  t=0\right)  .$}%
\end{figure}
Also, an \textquotedblleft extended\textquotedblright\ phase space trajectory
involving $X\left(  t\right)  $ and $dX\left(  t\right)  /dt$ gives some
improvement over the situation depicted in Figure 2, with the trajectory no
longer self-intersecting. \ However, the corresponding phase-space vector
field would still be singular with cusps appearing in $dX\left(  t\right)
/dt$ periodically in $X\left(  t\right)  $, similar to the cusps in the
potential $V\left(  X\right)  $. \ It is again possible to avoid the cusps by
considering the extended phase space trajectories as limits of positive energy
trajectories, but we leave that for the interested reader to pursue.

Finally, in terms of the coordinate transformation introduced in \cite{S},
namely, the angle
\begin{equation}
\theta=\arcsin\sqrt{x}\ ,
\end{equation}
the motion is completely unraveled, at least for $s=4$. \ In that case $E=0$
implies $\left(  \frac{d\theta\left(  t\right)  }{dt}\right)  ^{2}-\left(
\ln2\right)  ^{2}\theta^{2}\left(  t\right)  =0$, and the trajectory for all
$t$ is just that of a particle moving in a repulsive quadratic potential
(inverted SHO). \ This is the easiest example solvable in closed-form by the
Schr\"{o}der method \cite{CZ}. \ The exponentially growing solution,
\ $\theta\left(  t\right)  =2^{t}\theta\left(  0\right)  $, whose values in
the real-line cover of the circle are all physically distinct, immediately
leads to $x\left(  t\right)  $ for the particle moving through the various
$V_{P}\left(  x\right)  $, as is evident from (\ref{ExactTrajectory4}).

To amplify our discussion of the physics encoded in the switchback potentials,
we note that the chaotic behavior of the $s=4$ motion (i.e. the sensitive
dependence on initial conditions) follows entirely from the behavior of the
inverse Schr\"{o}der function. \ The sine function in (\ref{ExactTrajectory4})
discards any integer multiple of $2\pi$. \ So, for large and increasing times
the trajectory $\left.  x\left(  t\right)  \right\vert _{s=4}$ only depends on
digits farther to the right of the decimal in the numerical value of the
initial $\arcsin\sqrt{x}$, hence it is sensitive to those initial conditions.
\ Of course, this latter fact is well-known in the context of the discrete
$s=4$ map, and has been discussed time and again in the literature for the
case where $t$ is an integer (for example, see \cite{K}). \ As is evident from
(\ref{ExactTrajectory4})\ the discussion carries over \emph{mutatis mutandis}
to the continuous time case. \ The only point we add here is to emphasize that
it is $\Psi^{-1}$ which is responsible for the effect. \ Insofar as $\Psi
^{-1}$ follows from $\Psi$ and the latter follows directly from the potential
(as well as the other way round, as discussed in our Introduction) the chaotic
behavior follows directly from the specific multi-valued form of the potential.

For other values of $s$, from (\ref{Trajectory}) and (\ref{Interpolation}) we
again have $x\left(  t\right)  =\Psi^{-1}\left(  s^{t}\Psi\left(  x,s\right)
,s\right)  $, so the distinction between regular and chaotic behavior is again
determined entirely by the asymptotic behavior of $\Psi^{-1}$, hence the
asymptotic branch structure of $\Psi$, and hence the form of the switchback
potentials. \ The link between the analytic potential function and chaotic
behavior is arguably somewhat subtle, inasmuch as it requires some computation
to determine $\Psi$ and $\Psi^{-1}$, but the link is a direct one. \ Having
said that, there remains a case-by-case need for inspection of $\Psi^{-1}$ to
determine chaos or its absence. \ At this stage and in this paper, we do
\emph{not} purport to produce a universal analysis of all models --- we only
present methods that may facilitate such an analysis.

\section{Conclusion}

In summary, in this paper we\bigskip

--- Use a functional equation to convert nonlinear map equations into
dynamical systems in the continuous time domain, thereby facilitating the
study of such systems.

--- Illustrate this \emph{holographic} continuum interpolation technique with
the well-known model of the logistic map, explicitly for the chaotic case
$s=4$, an example which foreshadows an analysis for generic $s$.

--- Construct an intricate sequence of switchback Hamiltonians which drive the
system to just those phase-space trajectories resulting from such an analytic
interpolation of the logistic map, and produce the familiar chaotic behavior
for $s=4$.\bigskip

For nontrivial systems such as the chaotic $s=4$ logistic map illustrated
here, in order to produce the global features of the trajectories the
potential of the underlying Hamiltonian system \emph{must} change at each
switchback of the motion, as occurs when turning points are reached in finite
time. \ Therefore we have called these systems \emph{quasi-Hamiltonian}. \ The
trajectories dictate and completely determine the succession of switchback
potentials, in a richly extended analogy of inverse scattering techniques.

For values of $s$ between $2$ and $4$, a\emph{ }numerical analysis pursuant to
the techniques in this paper leads to quasi-Hamiltonian systems involving
switchbacks among sequences of potentials, eventually producing trajectories
that move between elaborate sets of turning points, in general. \ This
analysis is reported in full in a companion paper \cite{CV}. \ However, to
make the present paper more self-contained, and perhaps more convincing as to
the generality of the switchback effect, we have included some details for one
other case ($s=3$) in an Appendix.

To conclude, we encourage experimenters to look in various settings for the
particular type of continuous interpolates that we have described, including
fluid dynamical experiments, ecological systems, and, more generally, any
situation where an inherently continuous parameter (time, scale, etc.) has
been routinely sampled only at predetermined discrete values.

\begin{acknowledgments}
\textit{We thank D. Callaway for incisive questions and encouragement, D.
Sinclair for his emphasis on an appropriate title, and an anonymous reviewer
for suggestions to improve our presentation of the material. \ This work was
supported in part by NSF Award 0855386, and in part by the U.S. Department of
Energy, Division of High Energy Physics, under contract DE-AC02-06CH11357.}
\end{acknowledgments}

\section{Appendix: \ Some numerics for the $s=3$ case.}

The coefficients in the series expansion of $\Psi$ may be computed recursively
for any $s$, from (\ref{PsiSeriesCoefs}). \ While not immediately
recognizeable as the series expansion of any well-known function, nonetheless,
the coefficients may be used to estimate the radius of convergence of the
series (\ref{PsiSeries}) by the simple ratio test: \ $R_{\Psi}=\lim
_{n\rightarrow\infty}R_{\Psi}\left(  n\right)  $, with $R_{\Psi}\left(
n\right)  \equiv\left\vert d\left(  n-1\right)  /d\left(  n\right)
\right\vert $. \ To illustrate this consider the case $s=3$. \ For this case
numerical application of the ratio test leads to the data in Figure 4.%
\begin{figure}[ptb]%
\centering
\includegraphics[
trim=0.000000in 0.000000in 0.000000in 1.150459in,
height=2.1129in,
width=5.056in
]%
{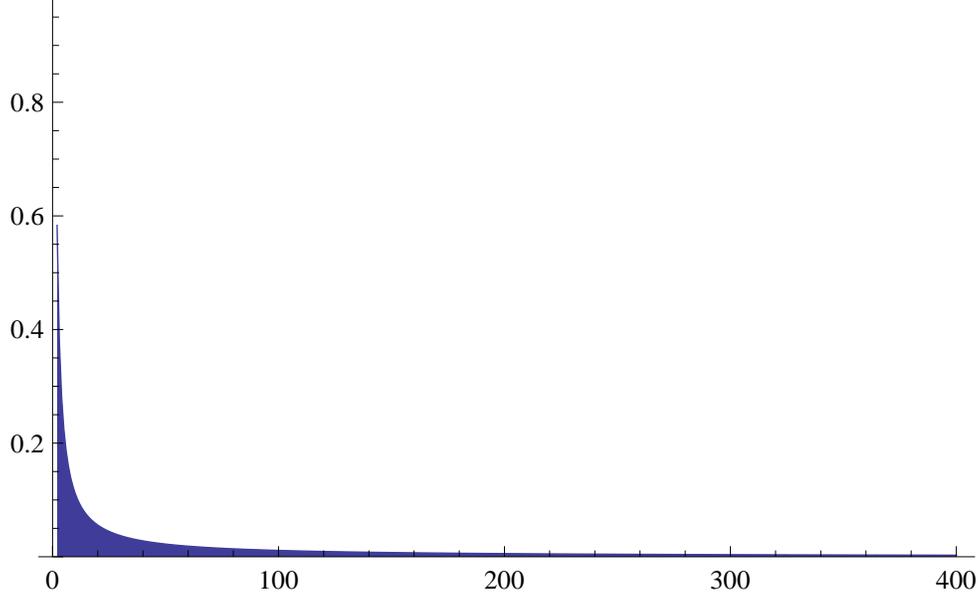}%
\caption{$R_{\Psi}\left(  n\right)  -s/4$ for $s=3$, plotted for $2\leq
n\leq400$, to show convergence towards $R_{\Psi}=3/4$.}%
\end{figure}
From the numerical data we infer that $R_{\Psi}=3/4=\left.  s/4\right\vert
_{s=3}$, as indicated earlier in (\ref{PsiSeriesRadius}). \ While this
numerical demonstration is not a proof of convergence, of course, we believe
it provides compelling evidence for convergence. \ 

Other values of $s$ produce numerical data qualitatively the same as the $s=3$
case, with only one noteworthy difference: \ While for $s>1$, the convergence
of the numerical data to the expressions in (\ref{PsiSeriesRadius}) is from
above, for $s<1$ the convergence is from below. \ 

A similar numerical calculation for the coefficients in the expansion of the
inverse function (\ref{PsiInverseSeries}) suggests $\Psi^{-1}\left(  x\right)
$ is an entire function of $x$, for $s=3$, much like the closed-form results
in (\ref{ExactCases}). \ That is to say, $R_{\Psi^{-1}}=\infty$ for $s=3.$

Combining the series data for $\Psi$, $\Psi^{-1}$, and $V$ with the functional
equations obeyed by them leads to an accurate numerical determination of the
functions and the switchback potentials, for any $s$. \ The methods are
described in detail in \cite{CV}. \ Numerical results for\ $\left.  \Psi
^{-1}\left(  \frac{1}{2}~s^{t}\right)  \right\vert _{s=3}$, with $-4\leq
t\leq9$, are shown in Figure 5 along with the expected asymptote, $\Psi_{\ast
}^{-1}=2/3$. \ This interval in $t$ corresponds to plotting $\Psi^{-1}\left(
x\right)  $ for $0\lesssim x\lesssim10^{4}$.%
\begin{figure}[ptb]%
\centering
\includegraphics[
height=3.4147in,
width=5.056in
]%
{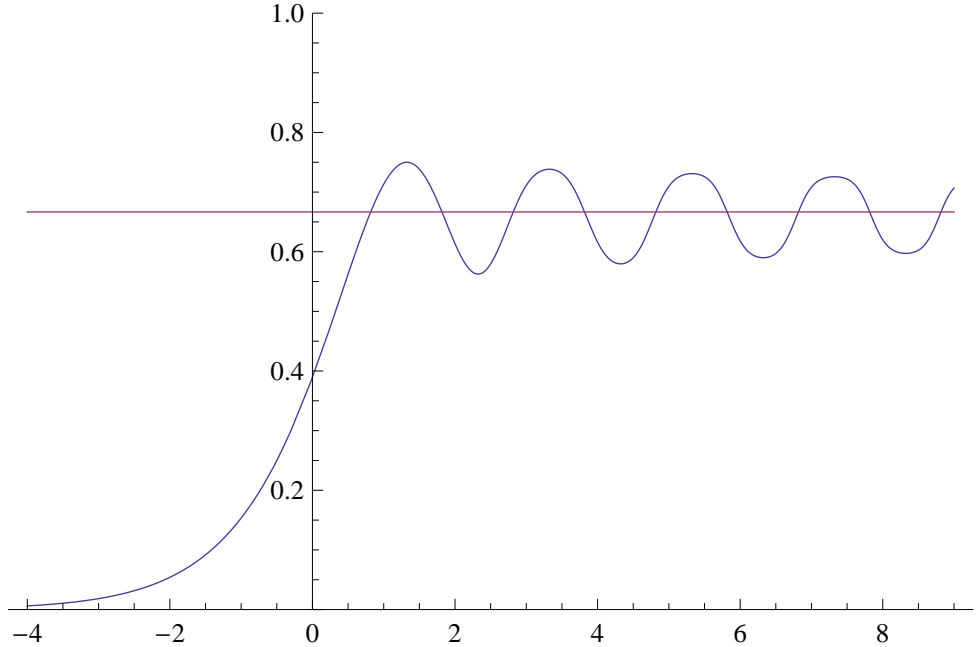}%
\caption{$\left.  \Psi^{-1}\left(  \frac{1}{2}~s^{t}\right)  \right\vert
_{s=3}$, in blue, for $-4\leq t\leq9$, from series plus functional methods.}%
\end{figure}

Numerical results for the first few potentials in the sequence, for $s=3$, are
shown in Figure 6.%
\begin{figure}[ptb]%
\centering
\includegraphics[
height=3.7011in,
width=5.9502in
]%
{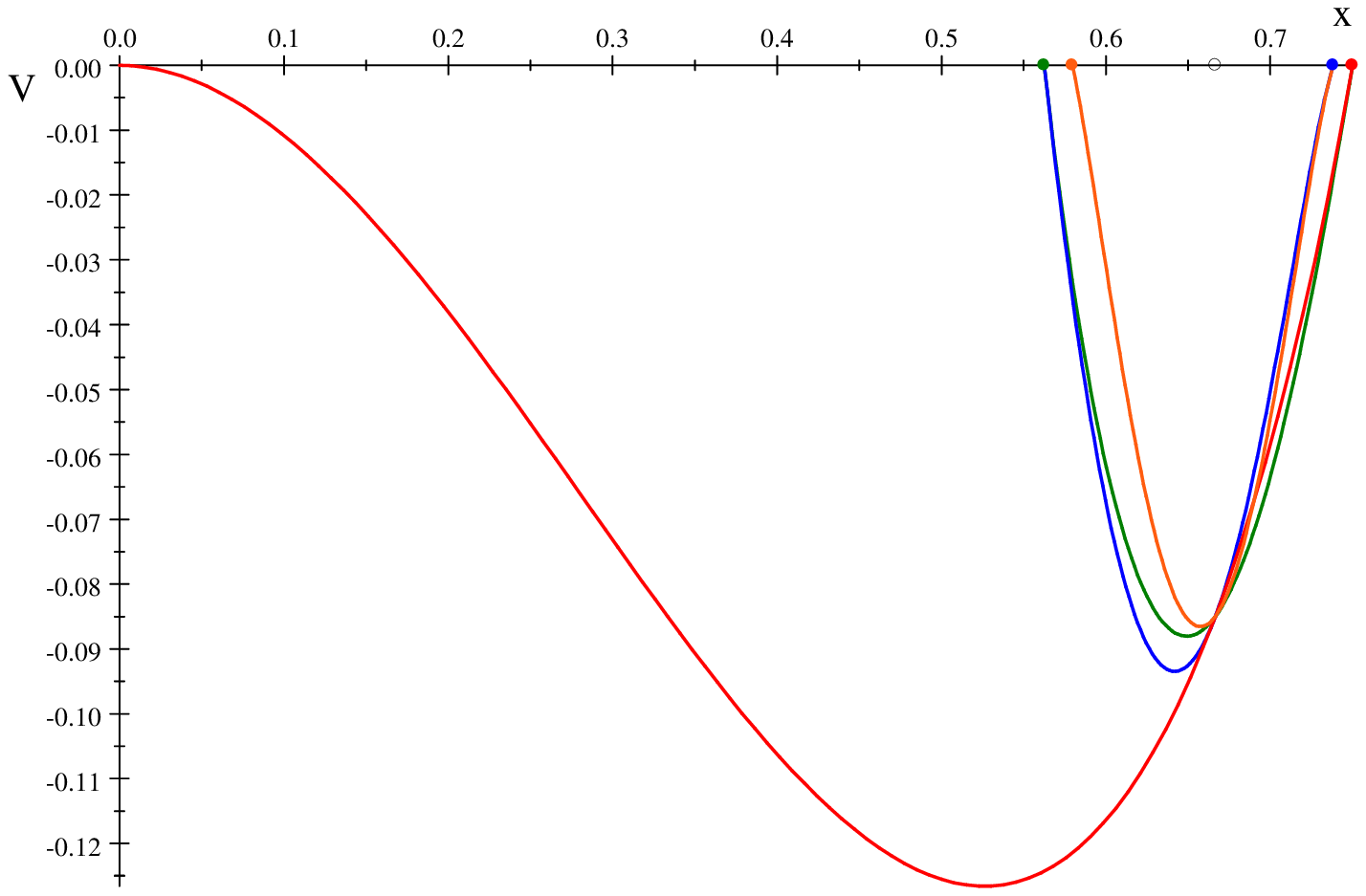}%
\caption{First four potentials in the sequence for $s=3$: \ $V_{0}$ red,
$V_{1}$ green, $V_{2}$ blue, and $V_{3}$\ orange.}%
\end{figure}
For this particular case, the sequence of potentials does not deepen
progressively, but rather the sequence converges, albeit very slowly, onto the
stable fixed point of the map, namely, $x_{\ast}=\left.  1-\frac{1}%
{s}\right\vert _{s=3}=2/3$, shown in the Figure as a small circle on the
$x$-axis. \ Hence the zero-energy particle trajectory will also converge onto
this fixed point, rigorously conserving $E=0$ in the process. \ 

Although the particle trajectory is not chaotic for $s=3$, nevertheless the
potentials are somewhat more exotic than the $s=4$ case in the sense that
there are an \emph{infinite} number of branch points for the $s=3$ case.
\ These may be obtained by iterating the action of the map on $x=1/2$. \ That
is to say, $\frac{1}{2}\rightarrow\frac{3}{4}=\allowbreak\frac{3}{2^{2}%
}\rightarrow\frac{9}{16}=\allowbreak\frac{3^{2}}{2^{4}}\rightarrow\frac
{189}{256}=\allowbreak3^{3}\frac{7}{2^{8}}\rightarrow\frac{37\,989}%
{65\,536}=\allowbreak3^{4}7\frac{67}{2^{16}}\rightarrow\cdots$. \ These branch
points are shown as the turning points of the potential sequence, $%
\begin{array}
[c]{ccccc}%
0.75, & 0.562\,5, & 0.738\,281\,, & 0.579\,666\,, & \cdots
\end{array}
$, represented by colored dots on the $x$-axis in Figure 6. \ The branch
points may also be seen by flipping the graph of $\Psi^{-1}$ in Figure 5 about
the SW-NE diagonal, in the usual way, to exhibit the various branches of
$\Psi$. \ 

This complicated branch structure for $s=3$ extends to other cases with
$2<s<4$ and most likely accounts not only for the failure to render them in
terms of simple, known functions, but also for some of the associated
wide-spread mystique that surrounds $\Psi$ and $\Psi^{-1}$ in these
cases.\ \ Of course, $s=2$ and $s=4$ only have branch points at $x=1/2$, and
at $x=0$ and $1$, respectively, as evident in the closed-form results
(\ref{ExactCases}) and as discussed in the text, so these cases are relatively
bland by comparison.


\begin{thebibliography}{99}
\bibitem {LogicalConnections}The results (\ref{Interpolation}) --
(\ref{Generalv}) establish strong ties between the $\Psi$-functional method
and the standard logical sequence conventionally used for time evolution,
wherein one begins with a given potential and then computes the trajectory.
\bibitem {Footnote1}Perhaps this is not without precedent. \ Consider a small
but macroscopic object made of weakly conducting dielectric material moving in
the electric field produced by some charged surfaces. \ Occasionally the
object might come in contact with one of the surfaces (i.e. hit a turning
point) and exchange some significant amount of electric charge in the process.
\ This would abruptly change the forces on the object, and correspondingly its
potential, affecting the object's motion and altering its subsequent
trajectory. \ The potential could change abruptly again if the object were to
come in contact with another charged surface. \ Of course, this process is
more difficult to model than that of the logistic map.
\bibitem {s=0}The $s=0$ case gives a simple, closed-form result, namely,
$\Psi\left(  x,s=0\right)  =x\left(  1-x\right)  $ and $\Psi^{-1}\left(
x,s=0\right)  =\frac{1}{2}\left(  1-\sqrt{1-4x}\right)  $. \ However, it is
also pathological in the sense it does not yield a particle trajectory from
(\ref{Interpolation}) as written. \ Of course, a \textquotedblleft scaling
limit\textquotedblright\ might be considered, with $\tau=t\ln s$ fixed as
$s\rightarrow0$. \ We leave this as an exercise for the reader.
\end{thebibliography}
\end{document}